 \documentclass[preprint,showpacs,preprintnumbers,amsmath,amssymb]{revtex4}
\usepackage{graphicx}
\usepackage{dcolumn}
\usepackage{bm}

\begin{document}

\title{Metastability Driven by Soft Quantum Fluctuation Modes}

\author{ Marco Zoli }
\affiliation{Istituto Nazionale Fisica della Materia -
Dipartimento di Fisica
\\ Universit\'a di Camerino, 62032, Italy. - marco.zoli@unicam.it}

\date{\today}

\begin{abstract}
The semiclassical Euclidean path integral method is applied to
compute the low temperature quantum decay rate for a particle
placed in the metastable minimum of a cubic potential in a {\it finite} time theory.
The classical path, which makes a saddle for the action, is derived in
terms of Jacobian elliptic functions whose periodicity establishes
the one-to-one correspondence between energy of the classical motion and temperature (inverse imaginary time) of the system. The quantum fluctuation contribution has been computed through the theory of the functional determinants for periodic boundary conditions. The decay rate shows a
peculiar temperature dependence mainly due to the softening of the
low lying quantum fluctuation eigenvalues. The latter are
determined by solving the Lam\`{e} equation which governs the
fluctuation spectrum around the time dependent classical bounce.

\end{abstract}


\maketitle

\section*{1. Introduction}

Quantum tunneling from a metastable state through a potential
barrier is a fundamental nonlinear phenomenon occuring in many
branches of the physical sciences \cite{landau,schulman,kramers,melnikov,hanggi,barone,larkvar,okuda,bao}.
The standard semiclassical approach to metastability has been
formulated since long by Langer \cite{langer} and Coleman
\cite{coleman} in the fields of statistical and nuclear physics
respectively. The main idea underlying such approach consists in
selecting a classical background which solves the Euler-Lagrange
equation and makes a saddle point for the action. Around the
background, the quantum fluctuations are treated in quadratic
approximation and their spectrum is obtained by solving a
Schr\"{o}dinger like stability equation whose potential is given
by the second spatial derivative of the metastable potential. The
semiclassical method finds a concise and powerful description in
the Euclidean path integral formalism \cite{feyn,fehi} in which the
time for the bounce to perform a full excursion (inside the classically
allowed region) is a measure of
the inverse temperature of the system. In the standard treatments
of metastability \cite{langer,coleman}, it is assumed that such
time is {\it infinite} and therefore the decay rate formula holds,
strictly speaking, only at $T=\,0$. For applications to specific
systems however, a precise knowledge of the decay rate at finite
$T$ (but within the quantum tunneling regime) may be of great
interest. To this purpose one has to build the {\it finite} time
theory of metastability for specific nonlinear potentials, setting
the crossover temperature between (low $T$) quantum and (high $T$)
activated regimes and find the shape of the decay rate when the
crossover is approached from below. Focussing on a widely
investigated model in nonlinear science, a particle in the one
dimensional cubic potential, I present in Section
2 the finite time solution of the Euler-Lagrange equation in terms
of the powerful Jacobian elliptic functions formalism \cite{wang}: this
generalizes the well known {\it infinite} time bounce which is
recovered asymptotically. I emphasize that the system is taken as
non dissipative therefore the {\it temperature} should
not be viewed here as a property of the heat bath
\cite{grabwei,rise,antunes}, but rather as a measure of the system
size along the time axis. Section 3 is devoted to the computation
of the classical action. Section 4 describes the method of the
semiclassical path integral and presents the calculation of the
overall quantum fluctuation contribution through the theory of the
functional determinants. Section 5 solves the stability equation
for the periodic potential defined by the classical background.
This permits to obtain analytically the lowest quantum fluctuation
eigenvalues as a function of the finite time/temperature. It is
shown in Section 6 that the softening of such eigenvalues close to
the crossover largely determines the peculiar shape of the decay
rate and its deviation from the prediction of the standard zero
$T$ theory. The conclusions are drawn in Section 7.

\section*{2. Cubic Potential Model}

\begin{figure}
\includegraphics[height=20cm,angle=0]{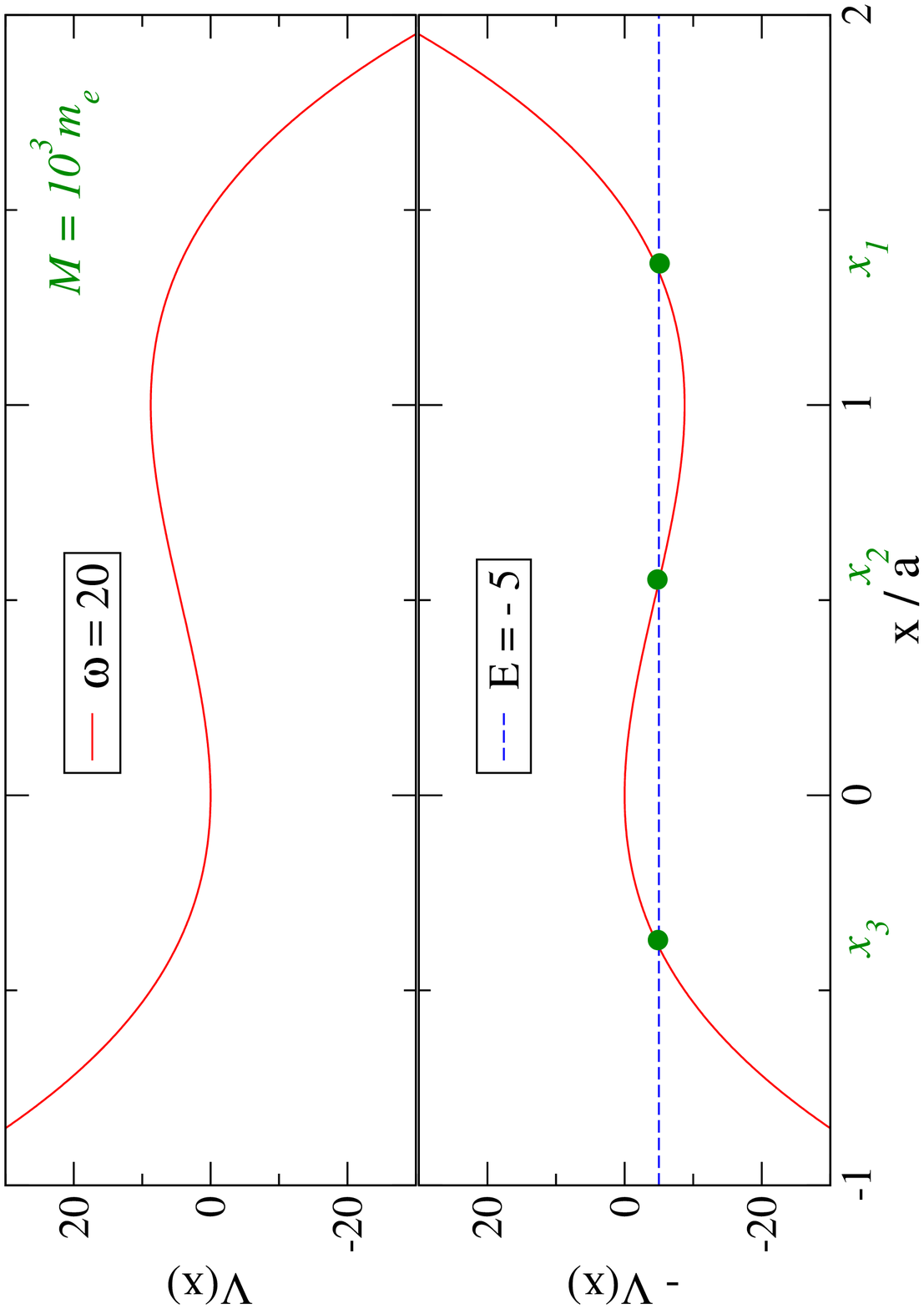}
\caption{\label{fig:1}(Color online) (a) Cubic potential in the
real time representation, (b) Cubic
potential in the imaginary time representation. $\omega$ is in $meV$. The
intersections with the constant energy $E$ (in $meV$) define the turning
points for the classical motion.}
\end{figure}

To begin, consider a particle of mass $M$ in the one dimensional
cubic anharmonic potential:

\begin{eqnarray}
V(x)=\, {{M\omega^2} \over 2}x^2 - {{\gamma} \over 3}x^3 \,\, ,
\label{eq:55}
\end{eqnarray}

plotted in Fig.~\ref{fig:1}(a) for $\hbar\omega=\,20meV$ and
$M=\,10^3m_e$, $m_e$ being the electron mass. Say $a$ the position
of the top of barrier whose height is $V(a)=\,\gamma a^3/6$ with
$\gamma=\,M\omega^2/a$. Let's take throughout the paper,
$a=\,1\AA$. At $x=\,0$ the particle is in a local minimum from
which it cannot escape classically. Thus, in the real time
formalism, the classical equation of motion admits only the
trivial solution $x_{cl}=\,0$. Physically however such local
minimum is metastable as quantum fluctuations allow the particle
to explore the abyss at $x \geq 3a/2$. In fact, a non trivial
classical solution can be found in the Euclidean space. Performing
a Wick rotation from the real to the imaginary time, $t
\rightarrow -i\tau$, the equation of motion reads:

\begin{eqnarray}
M\ddot{x}_{cl}(\tau)=\,V'(x_{cl}(\tau)) \,\, , \label{eq:1}
\end{eqnarray}

where $V'$ means derivative with respect to $x_{cl}$. In the
spirit of the semiclassical method, the particle path $x(\tau)$
has been split in the sum of a classical and a quantum component,
$x(\tau)=\,x_{cl}(\tau) + \eta(\tau)$. The Wick rotation is
equivalent to turn the potential upside down with respect to the
real time as shown in Fig.~\ref{fig:1}(b). Now it is clear that
the classical motion can take place in the reversed potential:
precisely, the particle moves back and forth between the turning
points $x_1$ and $x_2$ in which the particle velocity vanishes.

Integrating Eq.~(\ref{eq:1}), one gets:

\begin{eqnarray}
{M \over 2}\dot{x}_{cl}^2(\tau) - V(x_{cl}(\tau))=\, E \,\, ,
\label{eq:2}
\end{eqnarray}

with the constant $E$ representing the classical energy. Defining:

\begin{eqnarray} & &\chi_{cl}(\tau)=\,
{2 \over {3}}{{x_{cl}(\tau)} \over a} \, \nonumber
\\
& &\kappa=\,{{4E}\over {27 V(a)}} \,\, , \label{eq:56}
\end{eqnarray}

Eq.~(\ref{eq:2}) is easily integrated to yield:

\begin{eqnarray}
\tau - \tau_0 =\,\pm {{1 \over \omega}}
\int_{\chi_{cl}(\tau_0)}^{\chi_{cl}(\tau)} {{d\chi} \over
{\sqrt{-\chi^3 + \chi^2 + \kappa }}} \,\, , \label{eq:57}
\end{eqnarray}

where $\tau_0$ is the center of motion  between the turning
points. The boundary conditions on the classical motion define a
physical picture in which the particle starts from $x_2$ at the
time $\tau=\,-L/2$, reaches $x_1$ at $\tau=\,\tau_0$ and returns
to the initial position at $\tau=\,L/2$. Then, Eq.~(\ref{eq:57})
has a time reversal invariant solution whose period $L$ is finite
and dependent on $E$.

Looking at Fig.~\ref{fig:1}(b), one sees that the amplitude $x_1 -
x_2$ attains the largest value for the $E=\,0$ motion while $x_2$
and $x_3$ coincide. For $E < \,0$ motions, $x_3$ is negative. The
turning points $x_1$, $x_2$, $x_3$ are given by the zeros of the
equation $-\chi^3 + \chi^2 + \kappa=\,0$ ($\chi\equiv \, 2x/(3a)$)
which admits three real solutions for $\kappa \in [-4/27, 0]$,
that is for $E \in [-V(a),0]$. After some algebra I find:

\begin{eqnarray}
& &\chi_1=\, {1 \over {3}} + {2 \over {3}}\cos(\vartheta) \,
\nonumber
\\
& & \chi_2=\, {1 \over {3}} + {2 \over {3}}\cos(\vartheta -
2\pi/3) \, \nonumber
\\
& & \chi_3=\, {1 \over {3}} + {2 \over {3}}\cos(\vartheta -
4\pi/3) \, \nonumber
\\
& & \vartheta=\,{1 \over {3}}\arccos\bigl({{27 \kappa} \over {2}}
+ {1}\bigr) \,\, . \label{eq:58}
\end{eqnarray}

At the bounds of the energy range, Eq.~(\ref{eq:58}) yields:

\begin{eqnarray}
& &{\bf  E=\,0}  \Rightarrow  \chi_1=\,1 ;\,{}\, \chi_2=\,
\chi_3=\,0 \, \nonumber
\\
& & {\bf  E=\,-V(a)} \Rightarrow  \chi_1=\,\chi_2=\,2/3 ;\,{}\,
\chi_3=\,-1/3 \, \,\, . \nonumber
\\
\label{eq:59}
\end{eqnarray}

Thus, at the sphaleron energy $E_{sph}=\,|E|=\,V(a)$, the
amplitude of the finite time solution has to shrink into a point.
Let's find the general solution of Eq.~(\ref{eq:57}) by pinning
the center of motion at $\chi_{cl}(\tau_0)=\,\chi_1$ and using the
result \cite{grad}:

\begin{eqnarray}
& &\int_{\chi_{cl}(\tau)}^{\chi_{1}} {{d\chi} \over {\sqrt{(\chi_1
- \chi)(\chi - \chi_2)(\chi - \chi_3) }}}=\,{{2 F(\lambda,p)}
\over {\sqrt{\chi_1 - \chi_3}}} \, \nonumber
\\
& &\lambda=\,\arcsin\Biggl(\sqrt{{\chi_1 - \chi_{cl}(\tau)}\over
{\chi_1 - \chi_2}}\Biggr)\, \nonumber
\\
& &p=\, \sqrt{{\chi_1 - \chi_{2}}\over {\chi_1 - \chi_3}} \,\, ,
\label{eq:61}
\end{eqnarray}

where $F(\lambda,p)$ is the elliptic integral of the first kind
with amplitude $\lambda$ and modulus $p$.  Then, through
Eqs.~(\ref{eq:56}),~(\ref{eq:57}),~(\ref{eq:61}), I derive the
bounce solution of the {\it finite time} theory:

\begin{eqnarray}
& &x_{cl}(\tau)=\,{{3a} \over 2}\bigl[\chi_1 cn^2(\varpi,p) +
\chi_2 sn^2(\varpi,p)\bigr] \, \nonumber
\\
& &\varpi=\, \sqrt{{\chi_1 - \chi_3}}\,{\omega \over 2}(\tau -
\tau_0)\, \,\, , \nonumber
\\
\label{eq:62}
\end{eqnarray}

$sn(\varpi,p)$ and $cn(\varpi,p)$ are the {\it sine-} and {\it
cosine-} amplitudes respectively \cite{wang}. The modulus $p$ keeps tracks of
the classical mechanics through the second of Eq.~(\ref{eq:56})
and Eq.~(\ref{eq:58}).

At {\bf E =\,0, p=\,1}, the bounce of the {\it infinite time}
theory is recovered:

\begin{eqnarray}
& &x_{cl}(\tau)=\,{{3a} \over 2}cn^2(\varpi,1)=\,{{3a} \over 2}
sech^2\bigl({\omega \over 2}(\tau - \tau_0)\bigr) \, \,\, . \nonumber
\\
\label{eq:63}
\end{eqnarray}

At ${\bf E_{sph}, p=\,0}$,  the bounce solution is (as expected) a
point-like object set at the bottom of the valley in the reversed
potential: $x_{cl}(\tau)=\,a$. Thus, Eq.~(\ref{eq:62}) defines the
transition state which is a saddle for the action below the
sphaleron. Computation of Eq.~(\ref{eq:62}),  shows that the
bounce amplitude contracts by increasing the {\it energy over
potential height} ratio (in absolute value).  As the bounce is a
combination of squared Jacobi elliptic functions, its period is
$2K(p)$ with $K(p)=\,F(\pi/2,p)$ being the complete elliptic
integral of the first kind \cite{wang}. Hence, from
Eq.~(\ref{eq:62}), I get:

\begin{eqnarray}
\sqrt{{\chi_1 - \chi_3}}\,{\omega \over 4}L=\,K(p) \,\, , \label{eq:64}
\end{eqnarray}

which establishes the relation between the oscillation period and
the classical energy embedded in the turning points. As stated in
the Introduction, one can map the imaginary time onto the
temperature axis, $L=\,\hbar /(K_BT^*)$, where $T^*$ is the
temperature at which the particle makes the excursion to and from
the edge of the abyss for a given $E$.  Then, only periodic
bounces whose period is proportional to the inverse temperature
determine the decay rate and the {\it finite time} theory can be
viewed as a finite $T^*$ theory. From Eq.~(\ref{eq:64}) I get:

\begin{eqnarray}
K_BT^*=\,{{\hbar \omega} \over 4} {\sqrt{{\chi_1 - \chi_3}} \over
{ K(p)}} \,\, . \label{eq:64a}
\end{eqnarray}

\begin{figure}
\includegraphics[height=20cm,angle=0]{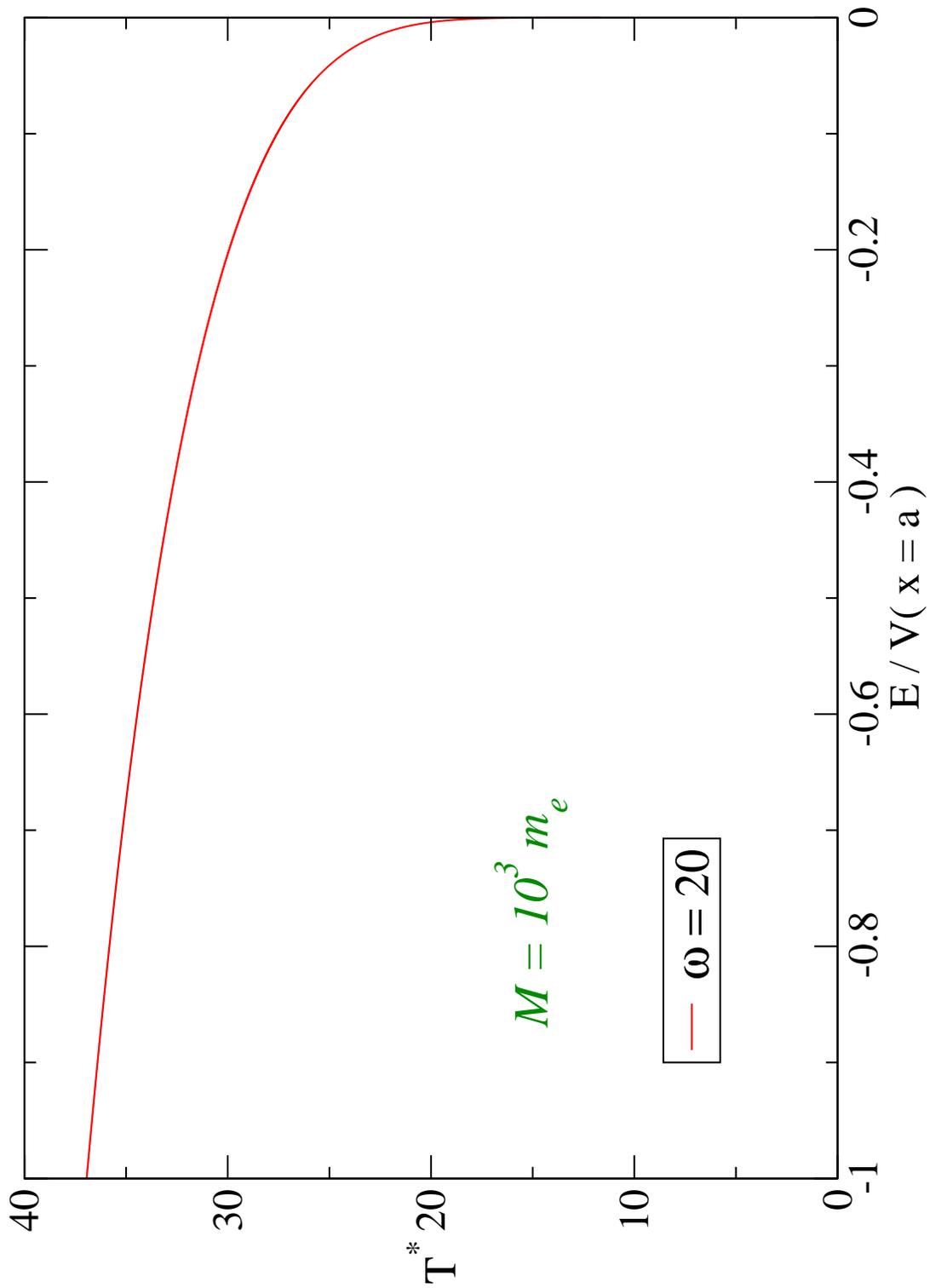}
\caption{\label{fig:2}(Color online)Plot of the Temperature (in
Kelvin) versus {\it Classical Energy over Potential Barrier
Height} ratio. $\omega$ is in meV.}
\end{figure}

Eq.~(\ref{eq:64a}) is plotted in Fig.~\ref{fig:2} on a linear
scale. Approaching $E=\,0$, $T^*$ consistently drops to zero while
the value at $E_{sph}$ defines the transition temperature $T^*_c$
between quantum and activated regimes.

Analytically, at the sphaleron, Eq.~(\ref{eq:64a}) yields

\begin{eqnarray}
K_BT_{c}^*=\,{{\hbar \omega} \over {2\pi}} \,\, , \label{eq:64b}
\end{eqnarray}

which represents the upper bound for the occurence of quantum
tunneling and precisely sets the Goldanskii criterion
\cite{gold,larkin} for a cubic anharmonic potential. Taking $\hbar
\omega=\,20meV$, I get $T_{c}^*=\,36.938K$. The following
calculations are carried out in the low temperature range up to
$T_{c}^*$.

\subsection*{3. Classical Action}

The classical action $A[x_{cl}]$ for the bounce in the finite time
theory can be computed in terms of the path velocity
$\dot{x}_{cl}(\tau)$ by the relations:

\begin{eqnarray}
& &A[x_{cl}]=\, M N^{-2} - E\cdot L(E) \, \nonumber
\\
& &{N}^{-2}=\, 2\int_{0}^{L/2}d\tau [\dot{x}_{cl}(\tau)]^2 \,
\nonumber
\\
& &\dot{x}_{cl}(\tau)=\,{{3a} \over 2}\mathcal{F}\cdot
sn(\varpi,p)cn(\varpi,p)dn(\varpi,p)\, \nonumber
\\
& &\mathcal{F}=\,-\omega (\chi_1 - \chi_2) \sqrt{{\chi_1 -
\chi_3}} \,\, , \label{eq:65}
\end{eqnarray}

where $dn(\varpi,p)$ is the {\it delta-} amplitude \cite{wang}.

Computation of Eq.~(\ref{eq:65}) requires knowledge of $L(E)$
through Eqs.~(\ref{eq:56}),~(\ref{eq:58}) and ~(\ref{eq:64}).

In the $E\rightarrow 0$ limit, from Eq.~(\ref{eq:65}), I get the
result

\begin{eqnarray}
{{A[x_{cl}]} \over \hbar} \rightarrow \,  {{6 M^3 \omega^5} \over
{5\hbar \gamma^2}} \,\, , \label{eq:66}
\end{eqnarray}

which serves as testbench for the computational method. The
dependence of the classical action on $1 / \gamma^2$ reflects the
well known fact that metastable systems are non perturbative and
provides the fundamental motivation for the semiclassical
treatment. Eq.~(\ref{eq:66}) permits to set the potential
parameters such as the condition ${A[x_{cl}] > \hbar}$ holds and
the semiclassical method is thus justified. As $M$ and $a$ have
been taken constant, ${{A[x_{cl}]}}\propto \omega$ in
the $E\rightarrow 0$ limit.

\begin{figure}
\includegraphics[height=20cm,angle=0]{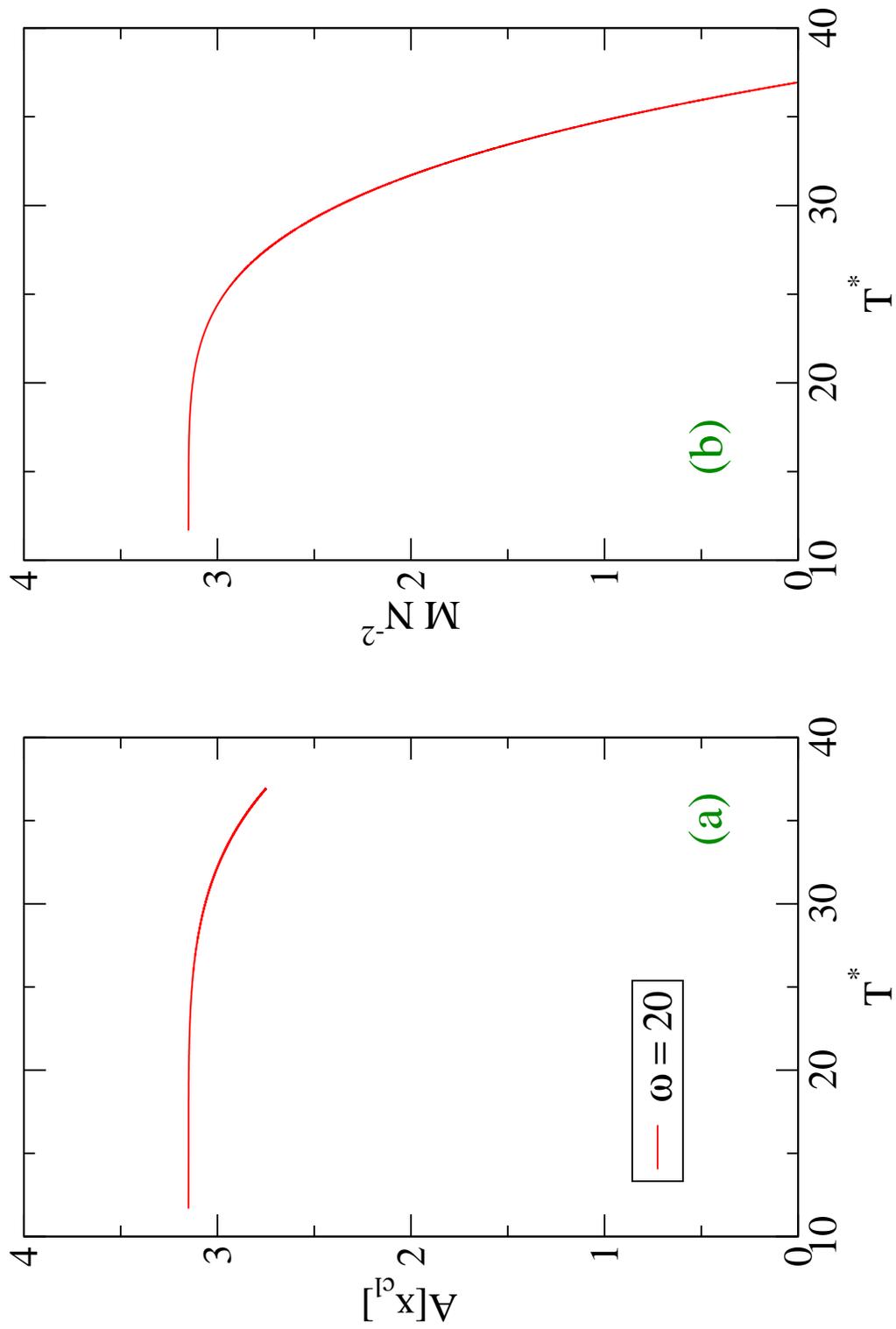}
\caption{\label{fig:3}(Color online) (a) Classical Action (in
units $\hbar$) versus Temperature; (b) Squared Norm of the bounce
velocity times particle Mass versus Temperature. $\omega$ is in
meV. }
\end{figure}

The classical action and the squared norm of the path velocity
(times $M$) are displayed in Fig.~\ref{fig:3}(a) and
Fig.~\ref{fig:3}(b) respectively. While, at low $T^*$, the two
plots are essentially identical the role of the term $E \cdot
L(E)$ in Eq.~(\ref{eq:65}) becomes more significant at increasing
$T^*$. At $T_{c}^*$, $N^{-2}$ vanishes whereas ${A[x_{cl}]}$ is
finite. Note that ${A[x_{cl}]}$ decreases smoothly versus $T^*$
confirming that the transition to the activated regime at
$T_{c}^*$ is of second order as suggested long ago
\cite{larkin,affl}.

In general, the criterion to establish the order of the
transitions in periodic tunneling systems has been formulated by
Chudnovsky \cite{chudno} through the behavior of the oscillation
period $L(E)$: {\it i)} a monotonic $L(E)$ below the sphaleron
implies $A[x_{cl}] < A_0$ for $T^* < T_c^*$, with $A_0$ being the
thermal action given by $A_0=\,\hbar V(a)/K_B T^*$.  At $T^*
=\,T_c^*$, both conditions $A[x_{cl}]=\, A_0$ and
$dA[x_{cl}]/dT=\, dA_0/dT$ are fulfilled hence the crossover from
the quantum to the thermal regime is expected to be smooth; {\it
ii)} on the other hand, a nonmonotonic behavior of $L(E)$ would
indicate a sharp transition. As it can deduced from
Fig.~\ref{fig:2}, $L(E)$ increases versus $E \in [- V(a), 0]$,
thus the case {\it i)} applies to the cubic potential in
Eq.~(\ref{eq:55}) and, consistently, the action is convex upwards
versus $T^*$. At the sphaleron, I find numerically:
$L(E_{sph})/\hbar=\,0.314meV^{-1}$. Note, from Eq.~(\ref{eq:64}), that such value
corresponds to $2\pi/\hbar\omega$ (being
$K(p=0)=\,\pi/2$) and this proves the correctness of the
computation. In fact at $E_{sph}$ the bounce is a point, that is a
static solution of Eq.~(\ref{eq:2}) but, near $E_{sph}$, the
periodic path is the sum of the sphaleron and a fluctuation with
negative eigenvalue $\varepsilon_{-1}$ whose period tends to
$L(E_{sph})=\,2\pi/\sqrt{|\varepsilon_{-1}|}$ \cite{park,blatter}.
Then one infers that, for $|E| \rightarrow E_{sph}$, the ground
state eigenvalue has to behave as: $\varepsilon_{-1}\rightarrow
-\omega^2$. This key point will be further investigated in Section
5.

\section*{4. Semiclassical Euclidean Path Integral}

This Section presents the calculation of the space-time Euclidean
path integral between the positions $x_i$ and $x_f$ connected in
the time $L$. In the semiclassical model and treating the quantum
fluctuations in quadratic approximation, the path integral reads:

\begin{eqnarray}
& &<x_f|x_i>_L=\,\exp\biggl[- {{A[x_{cl}]} \over {\hbar}} \biggr]
\cdot \int D\eta \exp\biggl[- {{A_f[\eta]} \over {\hbar}}
 \biggr] \, \nonumber
\\
& &A_f[\eta]=\,\int_{-L/2}^{L/2} d\tau \biggl({M \over 2}
\dot{\eta}^2(\tau) + {1 \over 2}V''(x_{cl}(\tau))\eta^2(\tau)
\biggr) \, \nonumber
\\
& &{{V''(x_{cl}(\tau))}}=\,{M\omega^2} \Bigl(1 - {2 \over a}
x_{cl}(\tau) \Bigr) \,\, . \label{eq:66+++}
\end{eqnarray}

Thus, to get the quantum fluctuation action $A_f[\eta]$, one has
to solve a second order differential problem which, after partial
integration in the first term, is formulated as follows:

\begin{eqnarray}
& &\hat{O} \eta_n(\tau)=\,\varepsilon_n \eta_n(\tau) \, \nonumber
\\
& &\hat{O}\equiv -\partial_{\tau}^2 + {{ V''(x_{cl}(\tau))}/
M}\,\nonumber
\\
& &\eta(\tau)=\,\sum_{n=\,-1}^{\infty} \varsigma_n \eta_n(\tau) \,\, ,
\label{eq:66++}
\end{eqnarray}

where the $\varepsilon_n$ are the quantum fluctuation eigenvalues
while the coefficients $\varsigma_n$ of the series expansion in
ortonormal components $\eta_n(\tau)$ define the measure of the
fluctuation paths integration in Eq.~(\ref{eq:66+++}):

\begin{eqnarray}
\int D\eta=\,\aleph \prod_{n=\,-1}^{\infty}
\int_{-\infty}^{\infty} {{d\varsigma_n}\over {\sqrt{2\pi\hbar/M}}}\,\, ,
\label{eq:66aa}
\end{eqnarray}

$\aleph$ depends only on the functional integral measure.

First, observe from Eq.~(\ref{eq:1}) that ${\dot{x}_{cl}(\tau)}$
satisfies the homogeneous equation associated to the second order
Schr\"{o}dinger-like differential operator $\hat{O}
\eta_n(\tau)=\,0$. This is a general consequence of the
$\tau$-translational invariance of the system. Hence,
${\dot{x}_{cl}(\tau)}$ is proportional to the ortonormal eigenmode
$\eta_0(\tau)$, ($\eta_0(\tau) \equiv \,N {\dot{x}_{cl}(\tau)}$)
with $\varepsilon_0=\,0$. The latter however cannot be the ground
state as the bounce solution (Eq.~(\ref{eq:62})) is non monotonic
and ${\dot{x}_{cl}(\tau)}$ has one node along the time axis within
the period $L$. This implies that the quantum fluctuation spectrum
has one negative eigenvalue corresponding to the ground state
\cite{i1}. Here lies the origin of metastability. Second, from
Eq.~(\ref{eq:65}), note that for any two points $\varpi_1,
\varpi_2$ such that $\varpi_2=\,\varpi_1 \pm 2K(p)$,
$\dot{x}_{cl}(\varpi_2)=\,\dot{x}_{cl}(\varpi_1)$. The important
consequence is that the fluctuation eigenmodes obey periodic
boundary conditions (PBC).

As $x_i$ and $x_f$ defined in Eq.~(\ref{eq:66+++}) coincide for
the periodic bounce, Eq.~(\ref{eq:66+++}) represents the single
bounce contribution $Z_1$ to the total partition function $Z_T$.
In fact, the latter also contains the effects of all multiple (non
interacting) excursions to and from the abyss which is equivalent
to sum  over an infinite number of single bounce contributions
like $Z_1$. Moreover, also the static solution of
Eq.~(\ref{eq:1}), $x_{cl}=\,0$ contributes to $Z_T$ by the
harmonic partition function $Z_h$ which can be easily determined
using the same measure in Eq.~(\ref{eq:66aa}). Summing up, $Z_T$
is given by

\begin{eqnarray}
& & Z_T=\,Z_h \exp(Z_1/Z_h)\, \nonumber
\\
& &Z_h=\,\aleph |Det[\hat{h}]|^{-1/2} \,\, , \label{eq:66a+++}
\end{eqnarray}

$Det[\hat{h}]$ ($\hat{h}\equiv \, -\partial^2_{\tau} + \omega^2$)
is the harmonic fluctuation determinant. Being the decay rate
$\Gamma$ proportional to the imaginary exponential argument
through the Feynman-Kac formula \cite{schulman}, it follows that
there is no need to determine $\aleph$ as it cancels out in the
ratio $Z_1/Z_h$.

Supposed to have solved Eq.~(\ref{eq:66++}), the quantum
fluctuation term in Eq.~(\ref{eq:66+++}) can be worked out by
carrying out Gaussian path integrals. Formally one gets:

\begin{eqnarray}
& &\int D\eta \exp\biggl[- {{A_f[\eta]} \over {\hbar}} \biggr]= \,
\aleph \cdot Det\Bigl[\hat{O}\Bigr]^{-1/2} \, \nonumber
\\
& & Det[\hat{O}]\equiv \, \prod_{n=\,-1}^{\infty}\varepsilon_n \,\, .
\label{eq:66++++}
\end{eqnarray}

The evaluation of Eq.~(\ref{eq:66++++}) is carried out through the
two following steps.

\subsection*{A. Zero Mode}

The Gaussian approximation leading to Eq.~(\ref{eq:66++++}) is
broken by the Goldstone mode arising from the fact that $\tau_0$,
the center of the bounce, can be located arbitrarily inside $L$.
The technique to overcome the obstacle is well known
\cite{larkin}: the divergent integral over the coordinate
$d\varsigma_0$ associated to the zero mode in the measure $D\eta$
is transformed into a $d\tau_0$ integral. Accordingly the
eigenvalue $\varepsilon_0=\,0$ is extracted from $Det[\hat{O}]$
and its contribution to Eq.~(\ref{eq:66++++}) is replaced as
follows

\begin{eqnarray}
(\varepsilon_0)^{-1/2} \rightarrow \sqrt{{{M {N}^{-2}}\over
{2\pi\hbar}}}L\, \,\, . \nonumber
\\
\label{eq:66+++++}
\end{eqnarray}

To be rigorous, this replacement holds in the approximation of
quadratic fluctuations \cite{kleinert} while higher order terms
may be significant around the crossover. It is also worth noticing
that Eq.~(\ref{eq:66+++++}) is often encountered in the form
$(\varepsilon_0)^{-1/2} \rightarrow \sqrt{{{A[x_{cl}]}\over
{2\pi\hbar}}}L$. However the latter is correct only in the low $T$
limit where $A[x_{cl}]$ equals $M {N}^{-2}$ (as made clear by
Fig.~\ref{fig:3}) while, approaching $T^*_c$, the difference
between the two objects gets large.  This fact is crucial in
establishing the behavior of the decay rate at the crossover as
shown in Section 6.

Thus, handled the zero mode, I turn to the evaluation of the
regularized determinant $Det^R[\hat{O}]$ defined by
$Det[\hat{O}]=\,\varepsilon_0 \cdot Det^R[\hat{O}]$.

\subsection*{B. Regularized Fluctuation Determinant}

The calculation of $Det^R[\hat{O}]$ is based on the theory of
functional determinants for second order differential operators
which was first developed for Dirichlet boundary conditions
\cite{gelfand} and then extended to general operators and boundary
conditions in several ways \cite{forman,tarlie,kirsten1}.

As a fundamental feature, to evaluate $Det^R[\hat{O}]$ one has to
know only the classical path which makes the action stationary. As
shown above the path velocity obeys PBC for any two points
$\varpi_1, \varpi_2$ separated by the period $2K(p)$. The latter
corresponds to the oscillation period $L$ along the $\tau$-axis.
It can be easily checked that also the path acceleration fulfills
the PBC. Then, the regularized determinant is given by
\cite{tarlie}:

\begin{eqnarray}
Det^R[\hat{O}]=\,{{<f_0 |f_0> \bigl(f_1(\varpi_2) -
f_1(\varpi_1)\bigr)} \over {{f_0(\varpi_1) W(f_0, f_1)}}}\,\, ,
\label{eq:67}
\end{eqnarray}

where $f_0, f_1$ are two independent solutions of the homogeneous
equation: $\hat{O} \eta_n(\tau)=\,0$. $f_0$ is obviously
$\dot{x}_{cl}$ while $f_1$ can be taken as:

\begin{eqnarray}
f_1=\,{{\partial {x}_{cl}} \over {\partial q}}\, ;{}\, q \equiv
\,p^2 \,\, , \label{eq:67+}
\end{eqnarray}

$W(f_0, f_1)$ is their Wronskian and $<f_0 |f_0> \equiv N^{-2}$ is
given by Eq.~(\ref{eq:65}).

The Wronskian, being constant along $\tau$, can be calculated in
any convenient point. Let's take $\tau_0$ as $f_0(\tau_0)=\,0$.
Then:

\begin{eqnarray}
& &W(f_0, f_1)\Bigr|_{\tau_0}=\,- \dot{f}_0(\tau_0)f_1(\tau_0) \,
\nonumber
\\
& &=\,{9 \over 8}a^2\omega^2 (\chi_1 - \chi_2){(\chi_1 - \chi_3)}
{{\partial \chi_1} \over {\partial q}} \, \,\, . \nonumber
\\
\label{eq:69}
\end{eqnarray}

Working out the calculation, ${Det^R[\hat{O}]}$ in
Eq.~(\ref{eq:67}) transforms into:

\begin{eqnarray}
{Det^R[\hat{O}]} =\,& &{{2} \over {\omega \sqrt{\chi_1 - \chi_3}
\bar{p}^2}} \Biggl[ {{E(\pi/2,p) - \bar{p}^2K(p)} \over {p^2}}
\Biggr]  \cdot {{<f_0 |f_0>} \over {W(f_0, f_1)}}\, \nonumber
\\  \bar{p}^2=\, & & 1 - p^2 \,\, ,
\label{eq:70}
\end{eqnarray}

which can be directly computed using
Eqs.~(\ref{eq:65}),~(\ref{eq:69}). $E(\pi/2,p)$ is the complete
elliptic integral of the second kind \cite{wang}.

It is however known in the theory of functional determinants
\cite{gelfand,kleinert} that only ratios of determinants are
meaningful in value and sign, such ratios arising naturally in the
path integral method as it has been pointed out above. In fact,
$Det^R[\hat{O}]$ would diverge in the $T^* \rightarrow 0$ limit
due to the fact that the determinant is the product over an
infinite number of eigenvalues with magnitude greater than one.
Consistently with Eq.~(\ref{eq:66a+++}), $Det^R[\hat{O}]$ has to
be normalized over $Det[\hat{h}]$  which, in the case of PBC, is:
${Det[\hat{h}]}=\,-4\sinh^2(\omega L/2)$ \cite{tarlie}. The
normalization cancels the exponential divergence and makes the
ratio finite.

Then, observing that for $E \rightarrow 0$ ( $T^* \rightarrow 0$):

\begin{eqnarray}
& &W(f_0, f_1)\Bigr|_{\tau_0}\rightarrow {9 \over 8}a^2\omega^2 (1
- p^2)\, \nonumber
\\
& &<f_0 |f_0>\, \rightarrow {6 \over 5}a^2\omega \, \nonumber
\\
& &K(p) \rightarrow \ln(4/\sqrt{1 - p^2}) \,\, , \label{eq:71}
\end{eqnarray}

from Eq.~(\ref{eq:70}), I finally get the finite ratio

\begin{eqnarray}
{{Det^R[\hat{O}]} \over {Det[\hat{h}]}} \rightarrow  -{{1} \over
{60\omega^2}} \,\, . \label{eq:71+}
\end{eqnarray}

The dimensionality $[\omega^{-2}]$ correctly accounts for the fact
that one eigenvalue has been extracted from $Det^R[\hat{O}]$. From
the computation of Eq.~(\ref{eq:70}), the following informations
can be extracted:

{\bf 1)} the $T^* \rightarrow 0$ limit given by Eq.~(\ref{eq:71+})
is in fact an excellent estimate up to $T^* \sim T^*_c/2$ whereas
a strong deviation is found at larger $T^*$ up to $\sim T^*_{c}$.

{\bf 2)} The ratio $Det^R[\hat{O}]/Det[\hat{h}]$ is negative for
any $T^*$ and this sign has physical meaning as it is due
precisely to the negative ground state eigenvalue
$\varepsilon_{-1}$ of the fluctuation spectrum. As $Z_1/Z_h$
contributes to the partition function by the square root of the
fluctuation (inverse) determinants ratio it follows that such
contribution is purely imaginary.

Moreover, close to the sphaleron, $T^* \sim T^*_{c}$, ${<f_0 |
f_0>} \propto p^4$ and $W(f_0, f_1) \propto p^2$, thus
$Det^R[\hat{O}]$ tends to zero as $Det^R[\hat{O}] \propto p^2$.
The fact that $Det^R[\hat{O}]$ vanishes at the crossover causes
the divergence of the inverse ratio. This divergence may look
surprising since the zero mode had been extracted from the
determinant: physically one realizes that there must be a quantum
fluctuation mode ($\varepsilon_{1}$) which softens by increasing
$T^*$ and ultimately vanishes at the sphaleron.

To understand in detail the key effects of the low lying
fluctuation eigenvalues $\varepsilon_{1}$ and $\varepsilon_{-1}$,
one has to determine them analytically by solving the stability
equation in Eq.~(\ref{eq:66++}). This is done in the next Section.

\subsection*{5. Lam\`{e} Equation}

Take Eq.~(\ref{eq:66++}) with the second derivative of the
potential given in Eq.~(\ref{eq:66+++}). Using Eq.~(\ref{eq:62})
and working out the algebra, I get the stability equation which
governs the fluctuation spectrum around the classical background:

\begin{eqnarray}
& &{{d^2} \over {d\varpi^2}} \eta_n(\tau) =\,\bigl[ l(l + 1)p^2
sn^2(\varpi,p) + \mathcal{A}_n \bigr] \eta_n(\tau) \, \nonumber
\\
& &\mathcal{A}_n=\,{{4(1 - 3\chi_1)} \over {\chi_1 - \chi_3}} -
{{4\varepsilon_n} \over {\omega^2(\chi_1 - \chi_3)}} \, \nonumber
\\
& &l(l + 1)\equiv 12 \,\, . \label{eq:73}
\end{eqnarray}

This is the Lam\`{e} equation in the Jacobian form for the case
$l=\,3$ \cite{whittaker}. For a given $l$ and $p$,
Eq.~(\ref{eq:73}) yields periodic solutions (which can be expanded
in infinite series) for an infinite sequence of characteristic
$\mathcal{A}_n$ values. The continuum of the fluctuation spectrum
stems from this sequence. However, being $l$ positive and integer,
the first $2l + 1$ solutions of Eq.~(\ref{eq:73}) are not infinite
series but polynomials in the Jacobi elliptic functions with real
period $2K(p)$ or $4K(p)$. Being the period of the potential,
$2K(p)$ plays the role of a lattice constant.

Then,  Eq.~(\ref{eq:73}) admits seven polynomial solutions with
eigenvalues $\mathcal{A}_n, n\in [-l,l]$, from which the
corresponding $\varepsilon_n$ are derived. However not all the
$\varepsilon_n$ are good fluctuation eigenvalues. In fact four out
of seven have to be discarded as their eigenfunctions do not
fulfill the PBC required for the fluctuation components:
$\eta_n(\varpi_1)=\,\eta_n(\varpi_1 \mp 2K(p))$. Thus, the three
good eigenmodes and relative eigenvalues in polynomial form are:

\begin{eqnarray}
& &\eta_{0}\propto \,sn(\varpi,p)cn(\varpi,p)dn(\varpi,p)
\nonumber
\\
& &\varepsilon_0=\,0 \, \nonumber
\\
& &\eta_1 \propto \,(sn^2\varpi - p^{-2})^{1/2}\biggl[sn^2\varpi +
{2 \over {p^2 + \mathcal{A}_1}}\biggr] \nonumber
\\
& & \varepsilon_1=\, \omega^2\Bigl(\alpha_1 -\alpha_2
\mathcal{A}_1 \Bigr) \nonumber
\\
& &\eta_{-1}\propto \,(sn^2\varpi - p^{-2})^{1/2}\biggl[sn^2\varpi
+ {2 \over {p^2 + \mathcal{A}_{-1}}}\biggr] \nonumber
\\
& & \varepsilon_{-1}=\,  \omega^2\Bigl(\alpha_1 -
\alpha_2\mathcal{A}_{-1} \Bigr)\, \nonumber
\\
& &\mathcal{A}_1= -(2 + 5p^2) - 2\sqrt{4p^4 - p^2 + 1}\, \nonumber
\\
& &\mathcal{A}_{-1}= -(2 + 5p^2) + 2\sqrt{4p^4 - p^2 + 1}\,
\nonumber
\\
& &\alpha_1\equiv\, 1 - 3\chi_1  \, \nonumber
\\
& &\alpha_2\equiv\, {{\chi_1 - \chi_3} \over 4} \,\, . \label{eq:75}
\end{eqnarray}

\begin{figure}
\includegraphics[height=20cm,angle=0]{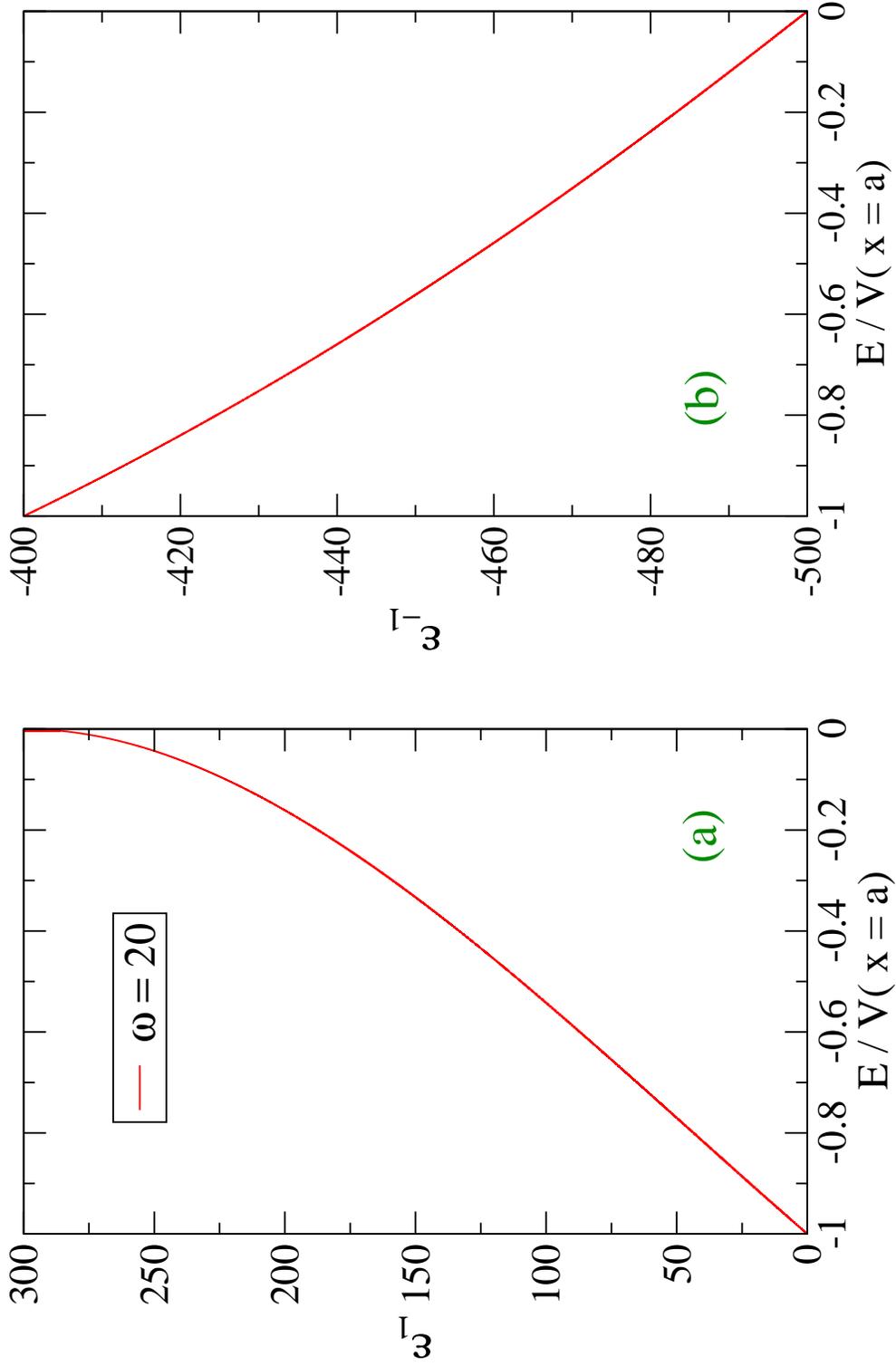}
\caption{\label{fig:4}(Color online) (a) Lowest lying positive and
(b) Ground State quantum fluctuation eigenvalues (in units
$\omega^2$) versus {\it Energy over Potential Barrier Height}
ratio. $\omega$ is in meV.}
\end{figure}

The plots of $\varepsilon_1$  and $\varepsilon_{-1}$ versus the
{\it energy over potential height} ratio are reported on
Fig.~\ref{fig:4}(a) and Fig.~\ref{fig:4}(b) respectively. Note
that:

{\bf i)} $\varepsilon_0=\,0$ is the zero mode eigenvalue correctly
recovered through the stability equation.

{\bf ii)} $\varepsilon_1$ lies in the continuum and, as it can be
easily deduced from Eq.~(\ref{eq:75}), it drops to zero close to
the sphaleron as $\varepsilon_1 \propto p^2$: this is precisely
the behavior previously envisaged for $Det^R[\hat{O}]$. Hence,
$\varepsilon_1$ is the soft mode driving the enhancement in the
decay rate below the sphaleron which is discussed in the next
Section. Observe that, for $p \rightarrow 0$, $K(p) \simeq \pi/2 +
\pi p^2/8$. Then, by Eq.~(\ref{eq:64a}), close to the crossover:
$\varepsilon_1 \propto T^*_c - T^*$. At $T_{c}^*$, $\varepsilon_1$
and $\varepsilon_0$ merge consistently with the double degeneracy
of the corresponding eigenmodes above the crossover.

{\bf iii)} $\varepsilon_{-1}$ is the negative eigenvalue
responsible for metastability. $\varepsilon_{-1}$ also softens (in
absolute value) with respect to the value $\varepsilon_{-1}=\,
-5\omega^2/4$ found at $E=\,0$. Interestingly, along the
temperature scale, the substantial reduction  starts up at $T^*
\sim T_{c}^*/2$, that is in the same range at which the classical
properties deviate from the predictions of the infinite time
theory. Finally, at the sphaleron, from Eq.~(\ref{eq:75}) I get
$\varepsilon_{-1}=\,-\omega^2$ thus confirming the prediction made
at the end of Section 3. This completes the analysis of the soft
eigenvalues which ultimately govern the quantum fluctuation
spectrum.

\section*{6. Decay Rate}

The decay rate $\Gamma$ of a metastable state is given in
semiclassical theory by

\begin{eqnarray}
\Gamma=\,A\exp(-B/\hbar)[1 + O(\hbar)] \,\, , \label{eq:100}
\end{eqnarray}

where $A$ and $B$ depend on the specific shape of the potential.
The investigation carried out so far allows us to identify the
coefficients $A$ and $B$ in Eq.~(\ref{eq:100}) with $\hbar
\sqrt{\bigl|Det[\hat{h}]/ Det[\hat{O}]\bigr|}/L$ and $A[x_{cl}]$
respectively. Then, the general expression for the finite
time/temperature $\Gamma(T^*)$ is:

\begin{eqnarray}
& & \Gamma(T^*)=\,\hbar  \sqrt{{{M {N}^{-2}}\over {2\pi\hbar}}}
\sqrt{\Biggl|{{Det[\hat{h}]} \over {Det^{R} [\hat{O}] }}\Biggr|}
\exp\biggl[-{{A[x_{cl}]} \over {\hbar}} \biggr]  \, \,\, . \nonumber
\\
\label{eq:76}
\end{eqnarray}

\begin{figure}
\includegraphics[height=20cm,angle=0]{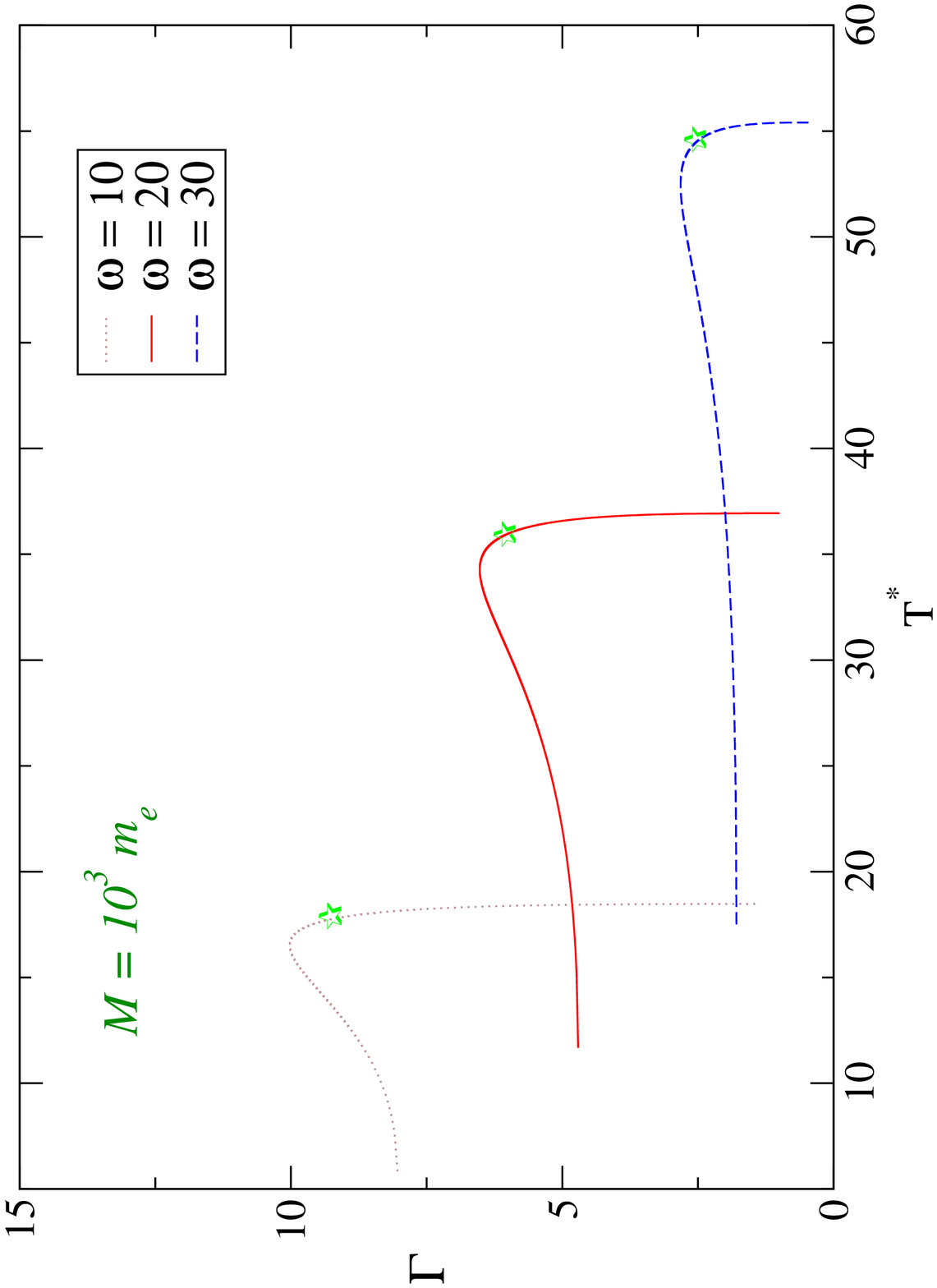}
\caption{\label{fig:5}(Color online) Decay rates (in $meV$) for a
particle in metastable cubic potentials (for three oscillator
energies in meV.) versus temperature up to $T^*_c$. The symbols
mark the temperatures $T^*_A$ in which the quantum decay rates
merge with the classical Arrhenius factors. }
\end{figure}

Eq.~(\ref{eq:76}) is plotted in Fig.~\ref{fig:5} against
temperature up to $T^*_c$ for three oscillator energies. While at
low $T^*$,  $\Gamma(T^*)$ merges with the constant decay rate of
the {\it infinite time} theory, an increase of $\Gamma(T^*)$ is
found in all plots above $T^* \sim T^*_c/2$ where the combined
effects of quantum fluctuations and classical action softening
become evident. Approaching the crossover, $\Gamma(T^*)$ deviates
from the  $T=\,0$ result  and reaches a peak value
$\Gamma(T_P^*)$ which is larger for lower $\omega$. This effect is
mainly ascribable to the soft eigenvalue $\varepsilon_1$. Note
however that for $\hbar \omega=\,10meV$,
$\Gamma(T_P^*)/\hbar\omega \sim 1$, signalling that the
application of the semiclassical method itself becomes
questionable. In fact, as noted below Eq.~(\ref{eq:66}), the
latter works when ${{A[x_{cl}]}
> {\hbar}}$ and such condition starts to be well fulfilled by the case
$\hbar \omega=\,20meV$ as shown in Fig.~\ref{fig:3}. Clearly the
$\omega$ values making the semiclassical method feasible also
depend on $M$ which has been assumed light in the present
discussion. Heavvier particle masses favor the condition
$\Gamma(T_P^*)/\hbar\omega < 1$ and sustain the applicability of
the semiclassical method over a broader range of $\omega$.

Above $T_P^*$ the decay rate smoothly merges with the classical
Arrhenius factor as ${{A[x_{cl}]}/ {\hbar}} \rightarrow
V(a)/K_BT^*_c$. The temperatures $T^*_A$, corresponding to the
symbols in Fig.~\ref{fig:5}, mark the effective values at which
quantum and thermal decay rates overlap. Beyond $T_A^*$ and
approaching $T^*_c$, the decay rate falls to zero as $\Gamma(T^*)
\propto (T^*_c - T^*)^{1/2}$ and the quantum tunneling ceases to
exist. The latter power law dependence is driven by
$\sqrt{N^{-2}/Det^R[\hat{O}]} \propto p$.

The increase found for the quantum decay rate up to $T_P^*$ and
the subsequent sharp drop is interesting also in view of a
comparison with activated systems described by classical
Ginzburg-Landau finite size models \cite{faris,tu} in which
spatio-temporal noise induces transitions between locally stable
states of a nonlinear potential \cite{chaud}. The changes in
radius and the stability conditions of metastable metallic
nanowires are an example of current interest
\cite{yanson,stafford}. In classical systems of finite size
$L$ a power-law divergence in the escape rate (with critical
exponent $  1/2$) is predicted once a critical lenghtscale $L_c$ is
approached at fixed $T$ \cite{stein1}. Instead, the quantum decay
rate of Eq.~(\ref{eq:76}) cannot be divergent as the small
parameter is $\hbar$ which, unlike the noise in classical systems,
cannot be varied as a function of $L$ at fixed $T$ (or
vice-versa) \cite{stein2}. Accordingly the quantum tunneling decay rate is small
and continuous.

Finally, it is worth pointing out that the decay rate may be
computed {\it independently} of the squared norm $N^{-2}$ as the
latter cancels out in Eq.~(\ref{eq:76}) by explicitly inserting
Eq.~(\ref{eq:67}). For this reason  the behavior of the decay rate
essentially depends on $Det^R[\hat{O}]/N^{-2}$ consistently with
the quadratic approximation for the quantum fluctuations which
enters the calculation at two stages: {\it a)} it determines the
form of the quantum action in Eq.~(\ref{eq:66+++}) and accordingly
leads to Eq.~(\ref{eq:66++++}); {\it b)} it allows us to replace
the inverse zero mode eigenvalue by the squared norm of the bounce
velocity, via Eq.~(\ref{eq:66+++++}).

\section*{7. Conclusion}

I have developed the finite time (temperature) semiclassical
theory for the quantum decay rate of a particle in the metastable
state of a cubic potential model. In the Euclidean path integral
formalism, the optimal escape trajectory emerges as the solution
of the Euler-Lagrange equation in terms of Jacobian elliptic
functions. Such solution is a time dependent bounce whose
periodicity naturally leads to relate the temperature $T^*$ to the
energy of the classical motion. Consistently one defines the
crossover temperature $T^*_c$ between quantum and activated
regimes which depends only on the fundamental
oscillator frequency $\omega$. As the path integral has been
solved treating the quantum fluctuations in quadratic
approximation, the calculations are confined to the low $\omega$
range, that is to the low temperature regime. In the numerical
analysis I have considered a light particle mass and
established, for this case, the lowest bound of $\omega$ values
which make the semiclassical method reliable.  The stumbling block
in the calculation of the quantum decay rate is the estimate of
the quantum fluctuation effect in the {\it finite} time theory. In
particular, I have {\it i)} derived a compact expression for the
overall fluctuation contribution to the path integral in terms of
the complete elliptic integrals and {\it ii)} solved the periodic
stability equation which yields the low lying fluctuation
eigenmodes and eigenvalues in polynomial form. The latter point
permits to quantify the softening of the lowest positive
and of the ground state (in absolute value) eigenvalues as $T^*_c$ is approached. The softening of the lowest positive eigenvalue is mainly responsible for the enhancement in the quantum
decay rate above the prediction of the infinite time (zero temperature) theory. The behavior of the decay rate has been studied in detail below $T^*_c$. At $T^* \sim T^*_c$, the thermal
activation sets in while the quantum decay rate drops
to zero according to the power law $\Gamma(T^*) \propto (T^*_c - T^*)^{1/2}$.
Similar conclusions may be drawn by the analyses of a quartic
metastable potential although the decay rate of the latter is
smaller than in a cubic potential having the same structural
parameters.

\begin{references}
\bibitem{i1}
In bistable $\phi^4$ potentials the classical path is monotonic
and the zero mode is the ground state of the fluctuation spectrum.
\end{references}
\end{document}